# Lifetimes of Fine Levels of Li Atom for 20 < n < 31 by Extended Ritz Formula

Muhammad Saeed, Zaheer Uddin

Department of Physics, University of Karachi, Karachi, Pakistan

Lithium and lithium-like elements look like hydrogen atoms if their two electrons and the nucleus are considered a core around which a single electron is orbiting. The energy and radii expressions for hydrogen atoms can be used for lithium and lithium-like elements; an important modification is introducing an effective principal quantum number. The effective principal quantum number differs from the principal quantum number of hydrogen by the quantum defect. Quantum defect has respective values for various levels of lithium and lithium ions. In this study, we used extended Ritz formulas to calculate quantum defects required to calculate energies of ns, np, nd, and nf series. Using these energies, we calculated transition probabilities and then the lifetimes of the lithium levels. The lifetimes were calculated with the published data; an excellent agreement was recorded. The work also extended the available list of lifetimes. Forty lifetimes are new and presented for the first time. a polynomial for each of the ns, np, nd, and nf series lifetimes has been produced that fits well the lifetime values

## 1 Introduction

C Laughlin et al., in 1973, employed the technique of nuclear-charge expansion for the calculation of multiplet strengths in lithium-like ions for $1s^2nl^2L - 1s^2ml'^2L'$ ($n, m \leq 4$) dipole transitions. They adopted two separate modes and used a modified screening approximation for rapid convergence of expansion. They finally predicted the values of oscillator strengths and radiative lifetimes for a series of atomic ions[1]. In 1979, K.T. Cheng et al. used the multiconfiguration Dirac-Fock method to calculate energy levels and wave functions. They calculated wavelengths, oscillator strengths, line strengths, and transition probabilities for $2s^n2p^m$ configurations for the E1, M1, and E2 transitions from Li–like to F-like ions [2]. In 1993, Martin et al. applied the quantum defect orbital and relativistic quantum defect orbital techniques for the calculation of the oscillator strengths for various transitions in Lithium atoms and Lithium-like ions (from z = 3 to z = 45);

the calculations were performed through standard dipole length operator. They obtained good results through relativistic quantum defect orbital techniques [3]. W. I. McAlexander et al. used an arrangement in which an ultracold $^6Li_2$ isotope was confined in a magneto-optical trap. Through photoionization, they recorded the spectra of the high-lying vibrational levels of $^6Li_2$ associated to $1^3\Sigma_g^+$ State which correlates to a $2P_{1/2}$ state and a $2S_{1/2}$ state atom. They used a model potential and obtained energy eigenvalues for the state $1^3\Sigma_g^+$. The eigenvalues were fitted to the experimentally measured vibrational levels and finally extracted the $2P$ lifetime equal to 26.99±0.16 ns [4]. Neng-Wu Zheng et al., in 2000, attempted to define the precise potential field in atoms. They proposed a new form of potential for the Weakest Bound Electron and derived the related wavefunctions. Consequently, the integral values of principal and orbital quantum numbers and the atomic number are replaced by non-integral values denoted as n*, l*, and Z*, respectively. Finally, they calculated the values of transition probabilities for the Li I and lithium-like Be, B, C [5]. N, and F. Çelik, G, in 2007, used two techniques: the WBEPMT and the Exact Quantum Defect (EQD) theory. He computed 108 transitions for single and multiple lines in Li I. The outcomes of both methods were compared with accepted values of (a) the NIST database for all transitions, (b) Opacity Project for multiplet transitions, and (c) MCHF for selective transitions from lower levels [6]. In 2017, A B Bichkov et al. used the method of ultrashort-pulse X-rays to evaluate the photoionization probability of Lithium atoms. The trajectory method was used to study the damaging dynamics. This technique allows the researchers to circumvent various restrictions posed by other techniques while analyzing the dynamics of the damaging sample procedure. They calculated the photoionization probabilities of various states of Li atom for field frequencies related to photon energies of 8 eV– 8 KeV and wavelengths from 1 to 1500 Å [7]. In 2022, Siddiq et al. calculated the Rydberg energies and transition probabilities of lithium atoms for transitions from np (n ≤ 15) to 2s, 3s, 4s, and 5s states. They employed the weakest bound electron potential model theory and generated new data of transition probabilities with reasonable accuracy [8]. In 2023, Miklos Ronto et al. developed a multi-state optimization approach in which the Pollak–Martinazzo energy lower bound was converged with a correlated Gaussian basis set, and the Lower bounds to the ground and first excited state energies for Li I and Be I were calculated. They claimed the most precise results of energy lower bound, showing the effectiveness of lower bounds for obtaining close approximations of atomic energies [9]. Wiese and Fuhr 2009 tabulated about 3600 transition probabilities for the H atom and its isotopes, He and Li atoms, and the Li II ion. They also listed scaling relations for He II and Li III ions. Due to the degenerate energy levels of $^1_1H$, $^2_1H$ and $^3_1H$ the tables contain average transition probabilities for the hydrogen atom and its isotopes [10]. Kostelecky and Nieto presented a potential function to obtain analytical wavefunctions, which were then used to determine the transition rates of some spectral lines of Li I and Na I [11]. Fischer et al. determined the values of energy levels, lifetimes, and transition rates using multiconfiguration Breit-Pauli for the lithium sequence up to Z=8 [12]. Wiese and Fuhr worked on hydrogen, helium, and lithium and tabulated their atomic transition [13]. Experimental work on calculations of lifetimes and transition rates of Li I is limited; the available experimental work reported only transition rates of 2p and 3p levels [14-15].

## 2 Theory

The most energetic and active electron in an atom is the one which can most easily be excited or ionized. This valence electron is most weakly bound to the atom and can be regarded as a Weakest Bound Electron (WBE). Solving the Schrödinger equation with effective Hamiltonian for lithium atom and separating it into the radial and angular parts, one gets the radial part as under,

$$\frac{d^2R}{dr^2} + \frac{2}{r}\frac{dR}{dr} + 2\left(\varepsilon - V(r_i) - \frac{l(l+1)}{r_i^2}\right) = 0 \quad (1)$$

Here, $V(r_i) = A/r_i - B/r_i^2$ is the model potential that the WB electron experiences. Moreover, $A = -z^*$ (the effective nuclear charge) and $2B = d(d+1) + 2dl$. The parameter d converts the integral quantum numbers $n$ and $l$ into effective non-integral $n^*$ and $l^*$ such that $l^* = l - \delta_n$, $n^* = n - \delta_n$. The term $\delta_n$ is the quantum defect that is given by extended Ritz formulism. Finally, the energy of the Weakest Bound Electron is $\varepsilon = \left[-Z^{*2}/2(n - \delta_n)^2\right]$. In some cases, the graph between $\delta_n$ and $1/(n - \delta_o)^2$ is not a straight line, the Ritz extended the formula in a series of $1/(n - \delta_o)^2$, and is given below

$$\delta_n = a_o + \frac{a_1}{(n - \delta_n)^2} + \frac{a_2}{(n - \delta_n)^4} + \frac{a_3}{(n - \delta_n)^6} \quad (2)$$

The solution of equation (1) will be the radial functions $R(r) = \frac{P(r)}{r}$, which can be expressed in terms of Associated Laguerre Polynomials, as

$$R = \left(\frac{2Z^*}{n^*}\right)^{l^*+\frac{3}{2}} \sqrt{\frac{(n^*-l^*-1)!}{2n^*\Gamma(n^*+l^*+1)}} \exp\left(-\frac{Z^*r}{n^*}\right) r^{l^*} L^{2l^*+1}_{n^*-l^*-1}\left(\frac{2Z^*r}{n^*}\right) \quad (3)$$

The expressions for transition probability $A_{fi}$, electric dipole line strength $S$ and lifetime $\tau_J$ are as under,

$$A_{fi} = 20261 \times 10^{-6} \frac{(E_f - E_i)^3}{2l_i + 1} S \quad (4)$$

$$S_{LS} = [J_f, J_i, L_f, L_i] \left(\begin{Bmatrix} L_f & S & J_f \\ J_i & 1 & L_i \end{Bmatrix} \begin{Bmatrix} L_f & l_f & L_c \\ 1 & L_i & l_i \end{Bmatrix} P^{(1)}_{l_i l_f}\right)^2 \quad (5)$$

We know that in lighter atoms, LS coupling dominates. In addition to two 6J symbols the $S_{LS}$ the term also contains a matrix element $P^{(1)}_{l_i l_f}$, given by,

$$P^{(1)}_{l_i l_f} = l_> <n_i, l_i|r|n_f, l_f> = l_> \int_0^\infty r^3 R_{n_i l_i} R_{n_f l_f} dr \quad (6)$$

$$\tau_J = \sum_{J'} \frac{1}{A_{JJ'}} \quad (7)$$

## 3  Results and Discussion

Although a huge amount of research has been carried out on the spectral properties of Lithium atoms, the data on lifetimes for atomic lithium is very meager, which does not match the volume of research conducted on the lithium atom. This research work has been envisioned to start from scratch and first calculate the line energies of atomic lithium, which could be further utilized to work out transition probabilities and the lifetimes of ns, np, nd, and nf series of lithium atoms.

### 3.1  Line Energies and Transition Probabilities of Atomic Lithium

In this work, the level energies for $1s^2ns$, $1s^2$ np, and $1s^2nd$ ($n \leq 30$) Rydberg states of Li I have been calculated using the extended Ritz formula, the quantum defects are calculated using first few energies and the coefficients of eq. (2) are found using least square fitting. The coefficients for ns, np, nd, and nf series are given in table 1.

Table 1: Coefficients of extended Ritz formulas

| Series | $a_o$ | $a_1$ | $a_2$ | $a_2$ |
|---|---|---|---|---|
| ns | 0.399151 | 0.033278 | -0.01633 | 0.030584 |
| np | 0.046952 | -0.01972 | -0.03169 | 0.056322 |
| nd | 0.046927 | -0.01852 | -0.04972 | 0.138454 |
| nf | 0.002311 | -0.03573 | 0.654353 | -4.47604 |

### 3.2 Lifetimes of ns, np, nd, and nf states

The calculated transition rates were used to calculate the lifetime of ns, np, nd, and nf levels up to n < 31. The transition rates given on the NIST [17] website were also used to calculate these series' lifetimes. This work on transition probabilities is also compared with the published work of Lindgard [18]and Theodosiou [19]. Lindgard's work consisted of transition rates up to n<13, whereas the work of Theodosiou consisted of n < 21; this work adds ten new lifetimes for 20 < n < 31, i. e., forty lifetimes of four series of Lithium atoms are new and presented for the first time. The lifetimes of ns, np and nd, nf are given in table 3 and 4, respectively.

Furthermore, a polynomial for each of the ns, np, nd, and nf series lifetimes has been produced that fits well the lifetime values and is expressed. A third-degree polynomial for ns and nd, a fourth-degree polynomial for np, and a fifth-degree polynomial for nf series of Li I represent lifetimes up to n < 31. The coefficients of these polynomials are given in table 2. The formulas work well, and the difference in calculated and fitted lifetimes is less than 0.1% in most cases. However, the difference is large for each series' first lifetime; it is 2%, 6%, 10%, and 2% in 3s, 2p, 3d, and 4f lifetimes. However, this difference is also in an acceptable range.

Table 2: The Coefficients of Polynomials for lifetimes for ns, np, nd, and nf series of Li I.

| Series | $a_o$ | $a_1$ | $a_2$ | $a_3$ | $a_4$ | $a_5$ |
|---|---|---|---|---|---|---|
| ns | 0.8354 | -0.7840 | 1.9914 | 7.9686 | | |
| np | -0.0002 | 2.7999 | 0.3717 | 50.4530 | -10.3780 | |
| nd | 0.4692 | 0.1974 | 1.2881 | 4.8471 | | |
| nf | 0.0001 | 0.0023 | 0.8971 | 2.3784 | -12.4550 | 27.1070 |

Table 3: The lifetimes of ns and np series of Lithium atoms and the corresponding values determined by NIST data, Lingard, and Theodosiou results.

| | Lifetimes $\tau_{ns}(ns)$ | | | | | Lifetimes $\tau_{np}(ns)$ | | | |
|---|---|---|---|---|---|---|---|---|---|
| $n$ | This Work | NIST | Lindgard | Theodosiou | $n$ | This Work | NIST | Lindgard | Theodosiou |
| 3 | 30.13 | 29.88 | 30.32 | 30.04 | 2 | 27.23 | 27.11 | 27.32 | 27.24 |
| 4 | 56.30 | 56.09 | 56.65 | 56.29 | 3 | 207.09 | 210.99 | 216.40 | 212.18 |
| 5 | 102.27 | 101.71 | 103.00 | 102.46 | 4 | 388.54 | 390.56 | 402.60 | 391.22 |
| 6 | 171.94 | 170.68 | 173.20 | 172.44 | 5 | 608.69 | 598.07 | 627.80 | 610.25 |
| 7 | 270.08 | 268.82 | 272.90 | 271.06 | 6 | 912.46 | 905.92 | 940.00 | 913.49 |
| 8 | 401.65 | 399.82 | 405.30 | 403.30 | 7 | 1319.56 | 1318.38 | 1358.90 | 1319.45 |
| 9 | 571.66 | | 578.60 | 574.76 | 8 | 1846.58 | 1848.06 | 1856.10 | 1848.24 |
| 10 | 785.13 | | 788.00 | 783.44 | 9 | 2509.80 | | 2587.50 | 2510.54 |
| 11 | 1047.10 | | 1061.40 | 1057.11 | 10 | 3325.53 | | 3260.40 | 3328.00 |
| 12 | 1362.60 | | 1374.10 | 1369.49 | 11 | 4310.20 | | 3448.90 | 4310.12 |
| 13 | 1736.67 | | | 1745.77 | 12 | 5480.27 | | 5683.50 | 5490.30 |
| 14 | 2174.36 | | | 2186.09 | 13 | 6852.28 | | | 6856.38 |
| 15 | 2680.71 | | | 2695.43 | 14 | 8442.82 | | | 8270.24 |
| 16 | 3260.76 | | | 3278.95 | 15 | 10268.49 | | | 10276.38 |
| 17 | 3919.56 | | | 3941.81 | 16 | 12345.96 | | | 12355.00 |
| 18 | 4662.15 | | | 4689.03 | 17 | 14691.90 | | | 14702.03 |
| 19 | 5493.57 | | | 5525.69 | 18 | 17323.01 | | | 17334.06 |
| 20 | 6418.86 | | | 6456.82 | 19 | 20256.03 | | | 20267.71 |
| 21 | 7443.06 | | | | 20 | 23507.68 | | | 23518.79 |
| 22 | 8571.23 | | | | 21 | 27094.74 | | | |
| 23 | 9808.39 | | | | 22 | 31033.99 | | | |
| 24 | 11159.59 | | | | 23 | 35342.24 | | | |
| 25 | 12629.87 | | | | 24 | 40036.28 | | | |
| 26 | 14224.26 | | | | 25 | 45132.96 | | | |
| 27 | 15947.80 | | | | 26 | 50649.13 | | | |
| 28 | 17805.52 | | | | 27 | 56601.45 | | | |
| 29 | 19801.52 | | | | 28 | 63006.14 | | | |
| 30 | 21926.27 | | | | 29 | 69876.49 | | | |
| | | | | | 30 | 76431.79 | | | |

Table 4: The lifetimes of the nd and nf series of Lithium atoms and the corresponding values determined by NIST data, Lingard, and Theodosiou results.

| | Lifetimes $\tau_{nd}(ns)$ | | | | | Lifetimes $\tau_{nf}(ns)$ | | | |
|---|---|---|---|---|---|---|---|---|---|
| n | This Work | NIST | Lindgard | Theodosiou | n | This Work | NIST | Lindgard | Theodosiou |
| 3 | 13.95 | 14.58 | 14.86 | 14.64 | 4 | 72.06 | 72.30 | 72.39 | 72.34 |
| 4 | 32.24 | 33.33 | 33.42 | 33.49 | 5 | 138.50 | 139.67 | 139.9 | 139.85 |
| 5 | 62.00 | 63.40 | 63.70 | 63.89 | 6 | 236.33 | | 239.1 | 239.04 |
| 6 | 106.06 | 108.34 | 108.20 | 108.63 | 7 | 371.71 | | 376.3 | 375.93 |
| 7 | 167.24 | 171.61 | 169.80 | 170.53 | 8 | 550.82 | | 556.1 | 556.59 |
| 8 | 248.38 | 254.20 | 251.20 | 252.39 | 9 | 779.90 | | 787.6 | 787.11 |
| 9 | 352.31 | | 355.90 | 357.02 | 10 | 1065.22 | | 1073.5 | 1073.59 |
| 10 | 481.86 | | 485.90 | 487.25 | 11 | 1413.08 | | 1419.7 | 1422.17 |
| 11 | 639.87 | | 648.20 | 656.66 | 12 | 1829.86 | | 1841.5 | 1840.29 |
| 12 | 829.17 | | 833.10 | 834.96 | 13 | 2322.00 | | | 2331.03 |
| 13 | 1052.60 | | | 1059.72 | 14 | 2896.01 | | | 2902.44 |
| 14 | 1313.01 | | | 1320.55 | 15 | 3558.50 | | | 3560.47 |
| 15 | 1613.21 | | | 1621.12 | 16 | 4316.21 | | | 4311.22 |
| 16 | 1956.06 | | | 1964.20 | 17 | 5175.98 | | | 5160.82 |
| 17 | 2344.38 | | | 2352.63 | 18 | 6144.80 | | | 6115.42 |
| 18 | 2781.02 | | | 2789.27 | 19 | 7229.83 | | | 7181.14 |
| 19 | 3268.81 | | | 3276.83 | 20 | 8438.42 | | | 8363.92 |
| 20 | 3810.58 | | | 3818.19 | 21 | 9778.11 | | | |
| 21 | 4409.18 | | | | 22 | 11256.70 | | | |
| 22 | 5067.43 | | | | 23 | 12882.23 | | | |
| 23 | 5788.17 | | | | 24 | 14663.05 | | | |
| 24 | 6574.24 | | | | 25 | 16607.84 | | | |
| 25 | 7428.47 | | | | 26 | 18725.61 | | | |
| 26 | 8353.69 | | | | 27 | 21025.82 | | | |
| 27 | 9352.73 | | | | 28 | 23518.31 | | | |
| 28 | 10428.45 | | | | 29 | 26212.72 | | | |
| 29 | 11583.27 | | | | 30 | 29091.27 | | | |
| 30 | 12803.19 | | | | | | | | |

## 4 Conclusion

The extended Ritz formulas calculate the quantum defects in a principal quantum number; it is a third-degree polynomial in $(n - \delta_o)^{-2}$. The quantum defects were fitted and used to calculate energies up to n < 31 levels. With the help of these energies, the transition rates were calculated for the levels belonging to the ns, np, nd, and nf series. Transition rates were compared with our previous work [8], and excellent agreement was found. With the help of these transition rates, the lifetimes of said series were calculated and compared with those calculated from NIST data, Lindgard [18], and Theodosiou's works. The lifetimes of ns, np, nd, and nf series up to n < 21 were calculated theoretically by Theodosiou [19]; we have extended it up to n < 31. Forty lifetimes, ten of each series ns, np, nd, and nf, are presented for the first time. A general polynomial has been developed by least square fitting, excellently representing the calculated lifetime values.